\documentclass[runningheads]{svmult}

\usepackage{makeidx}   
\usepackage{graphicx}  
\usepackage{subeqnar}  
\usepackage{multicol}  
\usepackage{physprbb}  
\makeindex             

\newcommand{\beq}{\begin{equation}}
\newcommand{\eeq}{\end{equation}}
\newcommand{\del}{\vec{\nabla}}
\newcommand{\dd}{\partial}        
\newcommand{\schwz}{ {\dd  \ln P\rho ^{-5/3} \over \dd Z}}
\newcommand{\schwR} { {\dd  \ln P\rho ^{-5/3} \over \dd R} }
\newcommand{\balbz}{ {\dd  \ln T \over \dd Z}}
\newcommand{\balbR} { {\dd  \ln T \over \dd R} }

\begin{document}
\title*{Numerical Simulations of MHD Turbulence\protect\newline
in Accretion Disks}
\toctitle{Numerical Simulations of MHD Turbulence
\protect\newline in Accretion Disks}
%
%
\titlerunning{Numerical Simulations of Accretion Disks}
%
\author{Steven A.~Balbus
\and John F. Hawley}
\authorrunning{S.~A.~Balbus \& J.~F.~Hawley}
%
%
\institute{Dept. of Astronomy, University of Virginia, Charlottesville,
VA 22901}

\maketitle              

\begin{abstract}
{We review numerical simulations of MHD turbulence. The last decade has
witnessed fundamental advances both in the technical capabilities of
direct numerical simulation, and in our understanding of key physical
processes.  Magnetic fields tap directly into the free energy sources
in a sufficiently ionized gas.  The result is that adverse angular
velocity and adverse temperature gradients, not the classical angular
momentum and entropy gradients, destabilize laminar and stratified
flow.  This has profound consequences for astrophysical accretion
flows, and has opened the door to a new era of numerical simulation
experiments.} \end{abstract}

\section{Introduction} 

Magnetized, differentially rotating plasmas are subject to a powerful
linear instability, whose maximum growth rate is a factor of $\sim
10^2$ per orbit (Balbus \& Hawley 1991).  This is the magnetorotational
instability, or MRI.  Despite the fact that planar Couette flow is
exquisitely sensitive to nonlinear disturbances and flow lamina quickly
breakdown into turbulence, decades of investigation have failed to find
any comparable mechanism in a nonmagnetized Keplerian disk.  The
problems is that local Coriolis forces are larger than the effective
disruptive inertial force caused by the presence of shear.  The result
is that nonacoustic disturbances respond in a wavelike manner in a
disk, whereas no such response is possible in nonrotating shear flow.
Instead, the displacements of perturbed fluid elements are (linearly)
unbounded, eventually becoming ensnared in the shear flow, feeding a
breakdown to turbulence.

The astrophysical significance of this is associated with accretion
sources.  By dint of angular momentum conservation, accretion onto
compact objects invariably involves differentially rotating flow with a
centrifugal barrier enshrouding the central mass.  The presence of a
magnetic field leads to the development of the MRI, which causes fluid
elements to lose their specific angular momentum, and leads to the
accretion process itself.  Happily, for the purposes of this conference,
many of the details of this process may studied via large scale
numerical simulations.

In this paper we will briefly review the history and contributions of
numerical MHD simulations of accretion disks.  Numerical investigations
of this problem began in earnest only a decade ago, and their impact
has been profound.  They have taught us not only {\em what} happens in
a highly complex accretion flow, by varying flow parameters we have
often learned why it happens as well.  As a recent example of this, we
shall discuss in some detail the structure of nonradiating flows
(associated with black hole accretion) that numerical simulations have
revealed in the last year or so.  The topic is important, fascinating,
and not entirely free of controversy.

The review proceeds along the following outline.  In \S 2, we review
the physics of the magnetorotational instability in its most general
form.  \S 3 discusses simulations that focus on a local patch of an
unstable Keplerian disk.  By limiting dynamical range, these
calculations may include more sophisticated physics.  In \S 4, global
MHD simulations are summarized.  Given their much larger dynamical
range requirements, it is not yet possible to treat the more complex
fluids amenable to a local analysis, but even these simple fluids may
be of astrophysical relevance.  There is ample evidence now for a
non-radiative flows at the Galactic Center, and there is reason to
believe this is not a special case.  We conclude with a summary in \S
5.

\section{The Magnetorotational Instability}
We begin with a review of the MRI (Balbus \& Hawley 1998).
The dynamics of this instability are very simple, involving only
the notion of magnetic tension in the presence of rotational
forces. 

\subsection {Formal Calculation}

The simplest case involves an axisymmetric disk threaded by a weak,
vertical magnetic field, $\vec{B} = B\vec{e_Z}$, where $\vec{e_Z}$
is a vector in the $Z$ direction.  (We are using a standard $(R,
\phi, Z)$ cylindrical coordinate system.) 
The undisturbed flow consists of fluid elements on
circular orbits.  We make a displacement
$\vec{\xi} = (\xi_R, \xi_\phi, 0)$ in the disk plane.  The
time and space dependence  of the displacement
is a simple plane wave,
\begin{equation}
\vec{\xi} = \vec{\xi'} \exp (ikZ - i \omega t),
\end {equation}
which defines the wavenumber $k$ and angular frequency $\omega$.
The wave propagates along the magnetic field line at the Alfv\'en
speed
$$
v_A^2 = {B^2\over 4 \pi\rho},
$$
and is a consequence of the restoring field line tension 
$-(kv_A)^2 \vec{\xi}$.  

We go into a frame rotating at the angular velocity $\Omega$ of a
fiducial orbit.  In this frame, in addition to the magnetic tension
force, we must add a Coriolis force $-2\vec{\Omega} \vec{\times}
d\vec{\xi}/dt$, and a centrifugal force $R\Omega^2 \vec{e_R}$.  The
latter is exactly balanced by the inward gravitational force just at the
location of the fiducial orbit, and the residual tidal force amounts to
$-\vec{\xi} d\Omega^2/d\ln R$.  The equations of motion for a fluid 
displacement are thus
\begin{equation}
{d^2\xi_R\over dt} - 2\Omega {d\xi_\phi\over dt} = -
\left[(kv_A)^2 + {d\Omega^2\over d\ln R} \right] \xi_R,
\end{equation}
\begin{equation}
{d^2\xi_\phi \over dt} + 2\Omega {d\xi_R\over dt} = -
(kv_A)^2\xi_R.
\end{equation}
The coefficients $\Omega$ and $d\Omega^2/d\ln R$ are now taken to
be constant, since our calculation is local.  
Note as well that pressure forces are unimportant since $\vec{k}\vec{\cdot}
\vec{\xi} = 0$.  

Plane wave solutions of these equations satisfy the dispersion
relation
\begin{equation}\label{disp}
\omega^4 - \omega^2[\kappa^2 + 2(kv_A)^2] + (kv_A)^2\left[
(kv_A)^2 + {d\Omega^2\over d\ln R} \right] = 0,
\end{equation}
where $\kappa$ is known as the epicyclic frequency
\begin{equation}
\kappa^2 = 4\Omega^2 + {d\Omega^2\over d\ln R} = {1\over R^3}
{d(R^4\Omega^2)\over dR},
\end{equation}
i.e., $\kappa^2$ is just proportional to the angular momentum gradient.
In the absence of a magnetic field, fluid displacements would oscillate
about their unperturbed circular orbit radius at a frequency $\kappa$.
The displacements can easily be calculated, and they appear in the
rotating frame as elliptical paths, with the elements moving in a
retrograde sense relative to the unperturbed circular orbits.  It is
these ``epicycles'' that give $\kappa$ its name.

The relation (\ref{disp})
is a quadratic equation in $\omega^2$, and it is straightforward matter
to show that if $d\Omega^2/dR < 0$, there are $\omega^2 < 0$ unstable modes.
By way of contrast, it is {\em angular momentum}, not
the angular velocity, that must decrease outwards for
instability in an unmagnetized disk.  This has major astrophysical
consequences, because disks in nature almost all have a specific
angular momentum profile that rises with increasing radius, but an angular
velocity profile that decreases in the same direction.
Ignoring even a highly subthermal field in a stability analysis is 
a grave error, leading to qualitatively incorrect conclusions.
Magnetized and unmagnetized fluids behave very, very differently.

The maximum growth rate of the instability is a quantity of interest,
\begin{equation}
|\omega_{max}| = {1\over 2} \left| d\Omega^2\over d\ln R\right|,
\end{equation}
which students of galactic structure will recognize as the Oort $A$
value.  In a Keplerian disk it amounts to $0.75 \Omega$, an enormous
growth rate.  It has been speculated that it is impossible for any
instability feeding off differential rotation to grow more rapidly
(Balbus \& Hawley 1992).

\subsection {Qualitative Description}

The equations of motion have precisely the same form as those
that emerge in a simple mechanical problem.  Imagine two masses
connected by a spring, in orbit about a central mass.  If
one replaces $(kv_A)^2$ by a spring constant, say $K$, our
systems are identical.  This gives an easy way to envision
the behavior of the instability, with magnetic tension acting
like a spring obeying Hooke's law.  

Denote the first mass as $M_1$, and assume it is on an orbit slightly
closer to the center than mass $M_2$, which orbits farther out.  $M_1$
travels slightly faster than $M_2$ if the angular velocity increases
inwards, and the spring tension pulls back on it.  On the other hand,
$M_2$ is is pulled forward in its orbital trek.  This means that there
is a positive torque pulling forward on $M_2$, and a retarding torque
pulling back on $M_1$.  Thus $M_1$ gains angular momentum, $M_2$ loses
angular momentum.  With additional angular momentum, $M_2$ (the outer
mass) moves to a more distant orbit, while $M_1$ (the inner mass),
moves to an orbit farther in.  This stretches the spring yet further,
the tension rises, and the process runs away.  This is the underlying
cause of the magnetorotational instability.  Magnetized angular
momentum transport is an intrinsically unstable process, because the
more fluid elements separate, the greater the rate of angular momentum
exchange that is driving the separation in the first place.

\subsection {General Stability Criteria}

It is possible to study the behavior of axisymmetric modes in
great generality.  By way of comparison, we first give the
results for an unmagnetized gas.  The local stability
of adiabatic perturbations is then governed by the H\o iland criteria
(e.g. Tassoul 1978):
\beq
\label{hoilad1}
-{3\over 5\rho}(\del P)\cdot\del\ln P\rho^{-5/3}
+ {1\over R^3} {\dd R^4\Omega^2\over \dd R} \ge 0,
\eeq
\beq\label{hoilad2}
\left( - {\dd P\over \dd z} \right) \, \left(
{\dd R^4 \Omega^2\over\dd R} \schwz - {\dd R^4\Omega^2\over\dd z}\schwR
\right) \ge 0,
\eeq
which allow for the presence of both vertical and radial gradients
in the host medium.   

In the presence of a magnetic field, all angular momentum gradients
are replaced by angular velocity gradients in the stability
criteria (Balbus 1995):
\beq\label{hoilb1}
-{3\over 5\rho}(\del P)\cdot\del\ln P\rho^{-5/3}
+ {\dd \Omega^2\over \dd \ln R} \ge 0,
\eeq
\beq\label{hoilb2}
\left( - {\dd P\over \dd z} \right) \, \left(
{\dd \Omega^2\over\dd R} \schwz - {\dd \Omega^2\over\dd z}\schwR
\right) \ge 0.
\eeq
This generalizes the result found in our simple example from the
previous section, that the angular velocity, not the angular
momentum, must increase outward when $\Omega = \Omega(R)$.
We find here that angular velocity gradients are quite generally
the proper stability discriminants in a magnetized gas.

\begin{figure}[t]
\begin{center}
\includegraphics[width=.9\textwidth]{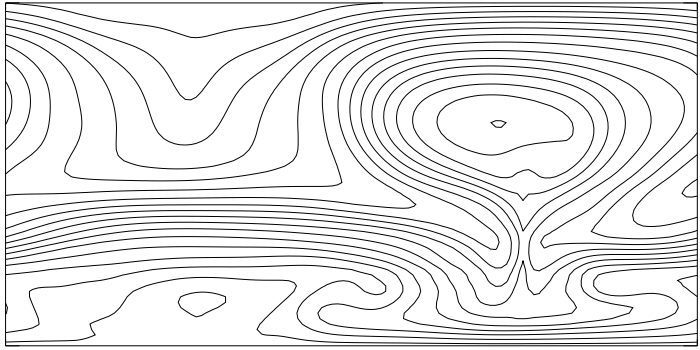}
\end{center}
\caption[]{
Development
of thermoclinic instability in a Schwarzschild-stable layer.
Magnetic lines of force are shown after one Alfv\'en crossing time,
initial seeding with rms 1\% random initial vertical velocity
perturbations.  Initial thermal energy density is 1600 times magnetic;
initial field lines are isothermal and horizontal; vertical grid runs from
$z=1$ to $2$, initial temperature profile is $1/z$, gravitational field
is $1/z^2$; $\chi$ is 0.05; grid is $128 \times 64$.  (From Balbus [2001],
simulation performed by J.~M.~Stone.)}
\label{therm}
\end{figure}

Finally we include the effect of a Coulomb conductivity.  
In the presence of a magnetic field, dilute astrophysical
plasmas conduct heat only along magnetic field lines, and then
of course only if there is a gradient along the field line.  
This is a regime of some relevance to black hole accretion
sources, in which the flow is characterized by very high
temperatures and low densities.  In this case, we find that
the stability criteria may be obtained from the previous simply
by changing the entropy gradients to temperature gradients
(Balbus 2001):
\beq\label{hoil1}
-{1\over \rho}(\del P)\cdot\del\ln T
+ {\dd \Omega^2\over \dd \ln R} \ge 0,
\eeq
\beq\label{hoil2}
\left( - {\dd P\over \dd z} \right) \, \left(
{\dd \Omega^2\over\dd R} \balbz - {\dd \Omega^2\over\dd z}\balbR
\right) \ge 0.
\eeq
Note that these requirements are independent of the thermal
conductivity coefficient.  This is a most surprising result:  a
nonrotating, adiabatic stratification of a temperature plus a small
sprinkling of magnetic field is highly unstable!  Figure (\ref{therm})
shows the results of a full nonlinear simulation of this instability,
as far as it could be followed.  The restriction that the heat flow
only along field lines is critical here, as a scalar thermal
conductivity (e.g. radiation) would serve only to stabilize
perturbations by dissipation.  Further discussion of this delicate
point may be found in Balbus (2001).

The three forms of the stability criteria show an ``evolution''
of replacing gradients of the extensive variables
$S$ and $L$ with gradients the intensive quantities $T$ and $\Omega$.
Changes in the former are energy sources,
\beq
dE = T\, dS + \Omega\, dL + ...,
\eeq
while changes in the latter are {\em free} energy sources,
\beq
d(E - TS -\Omega L) = -S\, dT - L\, d\Omega + ...,
\eeq
Gradients in the free energy generally mean that lower energy or higher
entropy equilibrium states are nearby.  When a dynamical path becomes
available to these states, instabilities are triggered.  Transitions
are swift: in all cases, the characteristic growth times are dynamical,
either a sound crossing or a shearing time.

\section {Local Nonlinear Simulations}

With the general stability conditions in hand, it is natural to ask
what are the nonlinear consequences of their violation?  In general the
answer is turbulent flow, and further progress depends upon numerical
simulation.

\subsection{Governing Equations}

Start with the fundamental $R$ and $\phi$ equations of motion:
\beq
\rho{\dd v_R\over \dd t} +\rho\vec{(v\cdot\nabla)} v_R -\rho{v_\phi^2\over R} = 
-\rho {\dd\Phi\over\dd R}
-{\dd P_{tot} \over \dd R} + \vec{(B\cdot\nabla)}
{B_R\over 4\pi} - {B_\phi^2\over 4\pi R},
\eeq
\beq
\rho{\dd v_\phi\over \dd t} +\rho\vec{(v\cdot\nabla)} v_\phi +\rho{v_\phi v_R\over R} =
-{1\over R} {\dd P_{tot}\over \dd \phi} + \vec{(B\cdot\nabla)}
{B_\phi\over 4\pi} + {B_\phi B_R\over 4\pi R},
\eeq
where 
$$
P_{tot} \equiv P + {B^2\over 8\pi}
$$
is the total gas plus magnetic pressure. 
To leading order, the disk is a simple Keplerian system, with angular velocity
\beq
R \Omega^2(R) = {GM\over R^2} \equiv {\dd\Phi\over \dd R}
\eeq
where $M$ is the central mass.  Thermal and magnetic forces are small
compared with the central gravitational force, though of course this 
``smallness scale'' is precisely the one we are concerned with!
We may express this quantitatively as
\beq
|(v_\phi -R\Omega)| \sim c_S \ {\rm (sound\ speed)} \sim v_A \ll R\Omega.
\eeq
Note that the azimuthally-averaged $\phi$ equation may be written
\beq\label{angmom}
\rho{\dd (R\rho v_\phi)\over \dd t} +\vec{\nabla}\cdot\left[\rho R\left(
v_\phi \vec{v} - v_{A\phi}\vec{v_{Ap}}\right)\right] = 0 
\eeq
where $\vec{v_{Ap}}$ is the poloidal Alfv\'en velocity, and the
all the angular momentum density and flux are understood to be 
averaged quantities.  The $\phi$-averaged
radial angular momentum flux may be read off directly
from equation (\ref{angmom}):
\beq\label{fjr}
{\cal F_J}_R = R \langle \rho(v_\phi v_R - v_{A\phi} v_{AR}) \rangle_\phi,
\eeq
which is an important quantity.

In the {\em local approximation}, the idea is to focus on a small patch
of the disk, fixing the new origin to corotate with disk fluid orbiting
at a particular
angular velocity, $\Omega_0$.  We measure all velocities relative
to $R\Omega_0$.  We ignore curvature effects in the local computational
patch of interest.  Formally, we work in the limit,
\beq
R\rightarrow \infty, \quad v_\phi \rightarrow \infty, \quad \omega\rightarrow
{\rm finite.}
\eeq
We define the velocity $\vec{w}$:
\beq
\vec{w} = \vec{v} - R\Omega_0\, \vec{e_\phi}.
\eeq
The value of $R$ at which $\Omega = \Omega_0$ will be denoted $R_0$.
In general we shall consider only small radial excursions 
from $R_0$,
\beq
R = R_0 + x, \quad  x \ll R_0.
\eeq
Thus,
\beq
{R} \Omega_0^2 - {\dd\Phi\over \dd R} = R (\Omega_0^2 - \Omega^2(R))\simeq
- x {d\Omega^2\over d\ln R}_0
\eeq
to leading order.  Local Cartesian coordinates can be defined by lining the $x$ and
$y$ axes along $R$ and $\phi$.  
In a frame rotating at $\Omega_0$, the local equations of motion for $w_X$ and $w_Y$
are 
\beq\label{start}
{\dd w_X\over \dd t} + \vec{w}\cdot\del{w_X} -2\Omega w_Y  + x
{d\Omega^2\over d\ln R} = -{1\over\rho}{\dd P_{tot} \over \dd x}
+ \vec{B}\cdot\del B_X
\eeq
\beq
{\dd w_Y\over \dd t} + \vec{w}\cdot\del{w_Y}
+2\Omega w_X = -{1\over\rho}{\dd P_{tot} \over \dd y}
+ \vec{B}\cdot\del B_Y
\eeq
The forms of the remaining dynamical equations remain unaffected by the
change to rotating coordinates.
They are the $Z$ equation of motion
\beq
{\dd w_Z\over \dd t} + \vec{w}\cdot\del{w_Z} = -{1\over\rho}{\dd P_{tot} \over \dd Z}
+ \vec{B}\cdot\del B_Z,
\eeq
the equation of mass conservation
\beq
{\dd\rho\over\dd t} + \vec{\nabla}\cdot (\rho\vec{w}) = 0 ,
\eeq
the internal energy equation,
\beq
(\gamma - 1) \rho \left(
{\dd\ \over \dd t} + \vec{w}\cdot\del\right)
{P\over\rho} = -P\del\cdot\vec{w}
\eeq
The induction equations for the magnetic field lose terms of relative
order $w/R\Omega$:
\beq\label{end}
{\dd B_i\over\dd t} +\vec{w}\cdot\del B_i = \vec{B}\cdot\del w_i
\eeq
where $i =X, Y, Z$.  

The set of equations (\ref{start})--(\ref{end}) completely describes
the local behavior of a magnetized accretion disk.  We must first,
however, specify the boundary conditions (BC) at the edges of the
computational domain.  The simplest approach is to use hard walls at
each of the radial, azimuthal, and vertical boundaries.  The first
nonlinear numerical investigations of local accretion disk behavior
were those of Hawley \& Balbus (1991), who used hard wall BC.  These
first simulations were highly restrictive, but sufficed to demonstrate
the existence of the weak field MRI, and confirm its analytic growth
rates and its independence of an azimuthal field.  These simulations
used a vertical field and more complex loop geometries as starting
configurations.  Because no periodic BC were needed, these early
simulations actually retained the curvature terms in the equations, so
that non WKB terms were present.  The detailed agreement between
analytic calculation and numerical results left no doubt whatsoever of
what was then a completely unexpected result: the combination of
Keplerian rotation and a weak magnetic field is extremely unstable.

Extended simulations require what are known as shearing-box BC.  In
this approach, the azimuthal boundary conditions are always periodic.
Vertical BC may be taken either as periodic or pure outflowing,
depending upon the problem on interest.  Because of the presence of
large scale shear, the radial BC are more complex (Hawley, Gammie, \&
Balbus 1995).  They may be described as ``quasi-periodic:''  the
computational domain is thought of as one brick in a wall extending to
infinity, each brick with the same fluid configuration as the next.
The velocity shear is continuous across the entire brick wall, so a
layer of bricks must slide with respect to the layer above and below!
As one layer slides over another, a fluid element leaving the
computational domain is replaced by its image on the opposite wall, but
not at the azimuth it has just vacated.  Instead it reappears at the
azimuth from the sliding brick in contact with the computational
domain.  The mathematical formulation of these BC may be found in
Balbus \& Hawley (1998).

\subsection {Local Axisymmetric Flow}

To run simulations over many local shear times, the shearing box
formalism must be implemented.  This was first done by Hawley \& Balbus
(1992) for two-dimensional axisymmetric flow.  A surprise emerged.
Though the evolution of an initial uniform vertical magnetic field was
in complete accord with analytic theory in the linear stages of its
development, the nonlinear stages hardly appeared turbulent at all.  In
fact, the linear stage seemed to continue indefinitely, with
exponentially growing streaming motions persisting.  This contrasted
sharply with the nonlinear behavior of a shearing box starting with a
vertical field whose mean value was zero---half upwards and half
downwards say, or sinusoidally varying.  In that case, turbulence
quickly developed after a few linear growth times, and then gradually
decayed, leaving no field at all at the end of the simulation!

This behavior can be understood with the help of Cowling's anti-dynamo
theorem (Moffatt 1978).  The theorem states that dynamo amplification is
impossible in an isolated, dissipative axisymmetric system.  To
understand what is meant by ``isolated,'' we work with the azimuthal
component of the vector potential, denoted $A$.  In axisymmetry, this
component alone suffices to determine the poloidal magnetic field.  The
mathematical heart of the theorem is that in the absence of resistance,
the integrated form of Faraday's induction equation may be manipulated
into the form 
\beq\label{cowl}
{\dd\ \over \dd t} \int A^2 \, dV = - \int A^2
\,\vec{v}\cdot\vec{dS} 
\eeq
where the left integral is over a volume containing the fluid (perhaps
infinite), and the right integral is over a bounding surface (perhaps
at infinity).  If periodic boundary conditions are used, or if $A^2$
falls off sufficiently rapidly, then the surface integral vanishes.
Hence, the average of $A^2$ remains constant for the fluid.  The
presence of any dissipation then causes an inevitable decline; there is
nothing to offset it.

When the mean vertical field vanishes, the vector potential satisfies
smooth periodic boundary conditions, and the anti-dynamo theorem
applies directly.  The magnetic field depends upon derivatives of $A$,
and its growth therefore requires a sort of continuous kneading of the
fluid, bringing different values of $A$ (which absent dissipation
is a fluid element label) ever closer together.  This continues on
smaller and smaller scales, but eventually the grid scale is hit, and
growth ceases at that point, with reconnection ensuing.  

The presence of a nonvanishing mean magnetic field implies that $A$ is
nonperiodic: it must have a component proportional to $x$.  Since
periodic boundary conditions no longer hold, the surface integral in
equation (\ref{cowl}) no longer vanishes, and poloidal field components
may grow at the expense of the free energy of differential rotation.
There is nothing unphysical about this set-up; real disks certainly can
be threaded by a magnetic field.  The question is do such disks really
exhibit the streaming behavior discussed above?

This was examined authoritatively by Goodman \& Xu (1994).  These
authors pointed out a remarkable fact: starting with a vertical field,
linear plane wave eigensolutions are in fact exact {\em nonlinear}
solutions to the equations of motion.  In the local approximation, the
gas really does appear to act like two orbiting masses connected by a
spring.  The existence and persistence in two dimensions of coherent
disk streams provided an explanation for a recurring puzzling behavior
seen in a number of earlier axisymmetric {\em global} MHD disk
simulations.  Uchida \& Shibata (1985) and Shibata \& Uchida (1986)
were interested in the creation of MHD jets and investigated this
problem by threading a disk of gas with a vertical magnetic field.  By
imparting less than the Keplerian value of the angular momentum to the
orbiting fluid elements, they hoped to duplicate the effects of slow
radial accretion---but without turbulence.  The infall produced radial
fields that became wrapped up by differential rotation into strong
toroidal fields, whose gradients in turn drove dynamical outflows along
the vertical field lines.  Some of these simulations, however, began
with a Keplerian disk embedded in a vertical magnetic field.  Such
disks also collapsed on a dynamic time scale.  At the time, the reason
for this was not at all clear.  This may now be understood as a global
manifestation of the streaming solutions studied by Goodman and Xu
(1994).

The question was whether in three dimensions the fluid behavior will
prove to be qualitatively different from the two dimensional case.
Will the streams remain stable in three dimensions?  Goodman \& Xu
noted that this ostensibly nonlinear question reduces to a {\em linear}
stability problem, but a linear stability problem perturbed about a
most unusual equilibrium solution.  The presence of vertically periodic
velocity streams renders the problem amenable to Floquet analysis
(Bender \& Orszag 1978), for which powerful mathematical techniques are
available.  The conclusion was that the new equilibrium of streaming
motions should be unstable.  The most important instability is a
magnetized Kelvin-Helmholtz mode, which appears for radial wavelengths
in excess of the streaming equilibrium flow's vertical wavelength.

\subsection{Local Three-Dimensional Simulations}

To determine the ultimate fate of the streams of the Goodman-Xu
solution or to study dynamo amplification requires implementation of
three-dimensional MHD codes.  By now, many shearing box studies have
been carried out, and a wide variety of models explored.  The simplest
consists of a homogeneous box, in which only the radial component of
the large scale gravitational field is retained, and the magnetic field
is initially uniform (Hawley et al. 1995; Matsumoto \& Tajima 1995).
The initial field geometry in these studies had some combination of
vertical and toroidal components.  A more complicated initial field
configuration, important for understanding dynamo activity, is to let
the initial field have a random character with vanishing mean (Hawley,
Gammie, \& Balbus 1996).  The presence of the vertical component of the
gravitational field produces a density stratification (Brandenburg et
al. 1995; Stone et al. 1996; Matsuzaki et al. 1997), which introduces
the possibility of magnetic buoyancy, a new effect.   Furthermore, by
bringing the pseudoscalar quantity $\Omega\cdot\del\rho$ into the
problem, stratification breaks chiral symmetry, i.e., the flow acquires
a ``handedness.''  This result is potentially important for the
development of local mean helicity in the turbulence, a feature upon
which much of classical kinematic dynamo theory is based (Moffatt
1978).

Stone et al.~(1996) carried out a series of such simulations spanning
two vertical scale heights in an initially isothermal disk.  With the
onset of the MRI, magnetic field rises out of the disc to establish a
highly magnetized corona.  The amplitude of the turbulence near the
midplane is determined more by local dissipation than by buoyant
losses, and the resulting stress levels are not greatly modified from
the nonstratified simulations.  The presence of a corona, however,
could have important observational consequences if  a significant
amount of dissipational heating occurs there.  Concerns about the
effects of the close-in vertical boundary conditions led Miller \&
Stone (2000) to carry out simulations with a larger computational
domain covering 5 scale heights above and below the equator.  They
demonstrated that indeed a strongly magnetized corona ($P_{\rm gas} \ll
P_{\rm mag}$) forms naturally from little more than a disk and a weak
seed field.

Three-dimensional studies also show the breakdown of the
two-dimensional streaming solutions (Hawley et al.~1995).  If the
computational box is large enough to allow an unstable radial
wavelength, streaming is disrupted within a few orbits, as the
Kelvin-Helmholtz instability noted by Goodman \& Xu (1994) leads to
fluid turbulence, which is the basis of significant outward angular
momentum transport.  This is because the turbulence is inherently
anisotropic; perturbations in the $x$ and $y$ components of the
magnetic field and the velocity are highly correlated.  Velocity power
spectra of the simulations are weighted toward the lowest wavenumbers,
i.e., there are significant fluctuations on scales comparable to the
computational domain size.  These large-scale fluctuations contain most
of the magnetic energy and contribute the lion's share of the stress.

Angular momentum is transported outward by the the MRI.  To understand what
is meant by that let
\beq
\vec{v}= R\Omega(R)\vec{e_\phi} + \vec{u}
\eeq
i.e., $\vec{u}$ is the velocity in excess of the local Keplerian
rotation.  (Note that $\vec{w}$ is by contrast the velocity in excess of the
local solid body rotation $R\Omega_0\vec{e_\phi}$.)  Then the radial angular
momentum flux from equation (\ref{fjr}) is
\beq
{\cal F_J}_R =   R^2\Omega \langle \rho u_R\rangle_\phi +
R \langle \rho ( u_R u_\phi - v_{AR} v_{A\phi} ) \rangle_\phi.
\eeq
The first term is simply the Keplerian angular momentum carried
directly by the mass accretion.  The second term is transport that is
present whether or not there is any accretion present.  This need not
imply that an individual fluid element is losing angular momentum
(though it generally is).  In its linear stage, an axisymmetric {\em
unmagnetized} convective instability transports angular momentum by
having low and high angular momentum fluid elements interpenetrate,
with no net mass flux.  This means an {\em inward} angular momentum
flux in a Keplerian disk, a result which is preserved in nonlinear
three-dimensional simulations (Stone \& Balbus 1996).

\subsection{Radiative Effects}

In general, thermal diffusion effects are not included in simulations.
If the object of study is a classical optically thick Keplerian disk,
this is a sensible approach to the dynamics.  Radiative diffusion
regulates the vertical temperature profile of the disk, but it does not
greatly influence dynamical stability.  

The linear stability of a magnetized, stratified, radiative gas was
recently addressed by Blaes \& Socrates (2001).  Despite the complexity
of the full problem, the MRI emerges at the end of the day unscathed,
its classical stability criterion $d\Omega^2/dR > 0$ remaining intact.
The maximum growth rate can be lowered however, particularly when the
azimuthal field approaches or exceed thermal strength.  (The same is
true for the ordinary MRI, as shown by Blaes \& Balbus [1994]).

The nonlinear problem has been carried through by Turner, Stone, \&
Sano (2002), who set up a local, radiative, axisymmetric, shearing box
flow.  The linear calculations of Blaes \& Socrates (2001) were
confirmed in detail, and the nonlinear flow was analyzed.  As in
standard MRI simulations, the stress is dominated by the Maxwell
component, which is a factor of a few larger than the Reynolds terms.
Photon diffusion plays a dual role in keeping the matter nearly
isothermal, and in creating over-dense clumps of gas of thermally
dominated gas.  The clumping occurs when radiation pressure support is
lost via diffusion on a dynamical time scale at large wave numbers.  In
this regime, radiative disks may be highly inhomogeneous.  At larger
optical depths and longer diffusion times, the disks look similar to
their nonradiative counterparts.

\subsection{Low Ionization Disks}  Protostellar disks, and possibly
cataclysmic variable disks, contain regions of low ionization
fraction---so low that the assumptions of ideal MHD break down.  Ohmic
dissipation becomes important, together with the Hall inductive terms.
(The latter arises because the distinction between the mean electron
and mean fluid velocities becomes important.)  The effect of Hall
electromotive forces on the MRI have been studied by Wardle (1999) and
by Balbus \& Terquem (2001).  Nonlinear numerical simulations including
ohmic resistivity have been done by Fleming, Stone , \& Hawley (2000);
Sano \& Stone (2002) have done simulations including both the ohmic and
Hall processes.  It can be shown that under rather general conditions,
if Ohmic dissipation is important, Hall electromotive forces are also
important (Balbus \& Terquem 2001; Sano \& Stone 2002).

The most interesting feature introduced by Hall electromotive forces is
helicity: the relative orientation of the angular velocity
$\vec{\Omega}$ and magnetic field $\vec{B}$ vectors matters (Wardle
1999; Balbus \& Terquem 2001).  The key point is that
$\vec{\Omega}\cdot \vec{B} > 0 $ configurations raise the maximum
growth rate in the presence of ohmic losses, and this aligned
configuration results in more vigorous transport in the local
axisymmetric simulations of Sano \& Stone (2002).  In configurations
where $\vec{\Omega}\cdot \vec{B}$ vanishes on average, the level of
turbulence is much more sensitive to the size of the ohmic dissipation
term.

In the simulations of Fleming et al. (2000) critical magnetic Reynolds
number $Re_M$ emerged below which turbulence is suppressed.  ($Re_M$ is
defined here as the ratio of the product of the scale height and sound
speed to the resistivity.)  When the mean field vanishes, $Re_M$ was
found to be surprisingly high, $\sim 10^4$.  In the presence of a mean
field, the critical $Re_M\sim 10^2$.  The interesting and important
question is whether the inclusion of Hall electromotive forces changes
these numbers by lowering them, i.e., making it easier to support
turbulence.  The Sano \& Stone (2002) axisymmetric simulations did not
reveal a large change, but questions on the maintenance of turbulence
more properly await a three-dimensional treatment.


\section{Global Disk Simulations}

Full global simulations are a demanding computational problem,
requiring extended evolutions at high resolution.  The fundamental
difficulties with global simulations are worth reiterating.  First one
has a severe problem with length scales.  The goal is to evolve
accretion disks from first principles, which means computing the
angular momentum transport self-consistently.  Angular momentum
transport is due to MHD turbulence, and the most-unstable MRI modes
will typically be much smaller than the disk pressure scale height
$H$.  One would also like to resolve the turbulent cascade through its
inertial range, if possible.  The disk itself extends from the black
hole horizon, $r_S$, out to  thousands of horizon radii.  Since all
time scales are more or less $\propto \Omega$, Kepler's law imposes a
difficulty as well.

One must be practical and work within well-chosen limitations.  It is
often helpful to forgo density stratification, for example.  One may
also restrict the dimensionality of the problem, by adopting
axisymmetry.  Finally, and most critically, one must make a judicious
choice of problem before embarking on a full global simulation.

\subsection{Two-Dimensional Simulations}

Axisymmetric MHD disk simulations have been used extensively in the
past, particularly to study jet formation processes.  Such simulations
date back at least to Uchida \& Shibata (1985).  It is only more
recently that two-dimensional simulations have directed primarily
at the internal dynamics of the disk itself and the resulting
accretion flow, rather than to the launching and collimation of jets.

From the astrophysical side, there has been considerable recent
interest in under-luminous accreting compact objects, inspired in no
small part by X-ray observations of the Galactic center (Melia \&
Falcke 2001).  Despite the inferred presence of a $2\times 10^6\,
M_\odot$ mass black hole, little of the expected radiation has been
detected by Chandra.  If the gas is able to radiate more rapidly than
it accretes, there seems little doubt that it forms a thin Keplerian
disk along the lines of Shakura \& Sunyaev (1973).  If the flow is
either too optically thick for the radiation to diffuse outward over an
accretion time, or too optically thin to radiate, then the fate of the
gas is much less clear.  Several ideas have circulated in the
literature, dating from the earliest days of accretion theory.  A small
sample:  Pringle \& Rees (1972), Ichimaru (1977), Begelman \& Meier
(1982), Rees et al. (1982), Abramowicz et al.  (1988), Narayan \& Yi
(1994, 1995), Narayan, Igumenschev, \& Abramowicz (2000); Abramowicz et
al. (2002).

From the numericist's perspective, it is a most welcome development
that there is evidence for, and intense interest in, nonradiating
flows.  For this is precisely the type of flow that is well-suited to
numerical simulation.  The first MHD simulations of nonradiating
accretion flows were carried out by Stone \& Pringle (2001).  Here,
accretion begins from an initial equilibrium torus, and is driven by
MHD stresses arising from the onset of the MRI.  The initial infall
phase is relatively smooth and dominated by the vertical field
``channel flow'' of the MRI (cf.~\S 3).  Subsequent evolution is
decidedly turbulent.  The resulting flow consists of an approximately
barotropic disk near the midplane, with constant $\Omega$ contours
parallel to cylindrical radii.   There was no strong correlation
between the specific angular momentum and entropy, as was found in
hydrodynamical simulations (Stone, Pringle, \& Begelman 1999).  A
substantial magnetized coronal outflow enveloped the disk.  Finally,
there was very little difference in the character of the solution if
dissipative losses were retained as heat or simply ignored.

While most of these features were destined to survive in
three-dimensional runs, two dimensions is generally a significant
limitation for the study of MHD turbulence Hawley (2000).  Poloidal
magnetic field cannot be indefinitely maintained in axisymmetry, by the
anti-dynamo theorem (e.g., Moffat 1978).  Indeed, toward the end of the
Stone \& Pringle (2001) simulations the turbulence clearly begins to
die away.  A  more minor concern is that axisymmetric simulations tend
to over-emphasize streaming modes, which produces coherent radial
magnetized flows rather than the more generic MHD turbulence.  Finally,
toroidal field instabilities cannot be simulated in axisymmetry.
Nothwithstanding these limitations,  by allowing a rapid investigative
turnover of plausible accretion scenarios, two-dimensional simulations
have proven to be a valuable tool, and a remarkably reliable one as
well.

\subsection{``Cylindrical Disks''}

The ``cylindrical disk'' is a global three-dimensional system allowing
for full radial and azimuthal dynamics, but ignoring the vertical
component of the central gravitational field.  Hence, there is no
vertical stratification.  Turbulence and magnetic field can be easily
sustained however, and far fewer vertical grid zones are required
compared with true three-dimensional simulations.

Armitage (1998) carried out the first such a calculation using the
standard ZEUS code and a grid covering the full $2\pi$ in azimuth,
running from $R=1$ to 4 with a reflecting inner boundary
and an outflow outer boundary, and covering a length of 0.8 in $z$.
Vertical boundary conditions were periodic.  The initial magnetic field
was vertical and proportional to $\sin(kR)/R$, with the sine function
argument linearly varying over $2\pi$ between $R=1.5$ and 3.5.  From
this initial state, turbulence rapidly developed along with significant
angular momentum transport.

A more extensive set of cylindrical simulations (Hawley 2001)
reaffirmed many of the conclusions of the local box models for
accretion disk turbulence driven by the magnetorotational instability,
as well as testing several technical aspects of global simulations.
Because the driving instability is local, a reduction in the azimuthal
computational domain to some fraction of $2\pi$ does not create large
qualitative differences.  Similarly, the choice of either an isothermal
or adiabatic equation of state has little impact on the initial
development of the turbulence.  Simulations that begin with vertical
fields have greater field amplification and higher ratios of stress to
magnetic pressure compared with those beginning with toroidal fields.

Recently, cylindrical disk calculations have been to the problem of the
star-disk boundary layer (Armitage 2002; Steinacker \& Papaloizou
2002).  Such simulations represent an important step forward, as
previous studies of boundary layers were modeled by a hydrodynamic
viscosity.  Significant dynamo activity was reported by both sets of
investigators.  In particular, Armitage (2002) finds an order of
magnitude larger field energy density in the boundary layer compared
with the average disk field.  Dissipative heating in the boundary
layer has yet to be simulated. 

\subsection{Three-Dimensional Simulations}

Global three-dimensional MHD disk simulations with full vertical
structure were presented by Hawley (2000).   These models started with
equilibrium tori containing either a weak poloidal or a weak toroidal
magnetic field. A torus is a useful initial condition for global
simulations because it can be well-resolved and wholly contained within
the grid.  The thickness of the torus depends upon the angular momentum
distribution, with constant specific angular momentum $l$ associated
with the thickest tori.  Such a structures may actually be generic in
active galactic nuclei, where the feed an an inner disk (Krolik 1999).
As the system evolution proceeds, the MRI develops rapidly.  Stresses
are dominated by the Maxwell component (a factor of several larger than
the Reynolds stress), which immediately redistribute angular momentum
from the initial non-Keplerian profile to nearly Keplerian state.  At
later times, the disks show rapid time variability, tightly wrapped,
low-$m$ spiral structure, and significant internal stress.

Machida, Hayashi, \& Matsumoto (2000) studied the evolution of a
constant angular momentum torus containing toroidal magnetic field.
Turbulence soon develops and magnetically dominated ($\beta < 1$)
filaments form.  The buoyancy of such fields leads to the development
of a strongly magnetized corona.

Steinacker \& Henning (2001) have recently revisited the question of
the influence of a large-scale vertical field on an accretion disk,
previously studied in axisymmetry.  They found strong accretion in the
disk (the authors described it as a collapse) driven by the MRI.
Coupling to a magnetized corona drives outflows, with the degree of
collimation depending on the strength of the field.

The nature of the magnetic stress near the marginally stable orbit of
black hole ($6GM/c^2$ for a Schwarzschild hole) has received renewed
attention.  This disk hydrodynamic treatments using a turbulent
viscosity lead to a vanishing stress in this region (Abramowicz \& Kato
1989), but matters can be more complicated when magnetic stresses are
involved (e.g. Novikov \& Thorne 1973; Gammie 1999; Algol \& Krolik
2000).  The question of what happens to the stress is of interest
because of the possibility of extracting energy from the hole's
rotation and communicating it to the disk.  It can be addressed via
three-dimensional MHD simulations.

Hawley \& Krolik (2001, 2002) investigated the evolution of a
magnetized accretion torus lying near the marginally stable orbit in a
pseudo-Newtonian model potential for the black hole
(Paczy\'nski \& Wiita 1980).   The work focused on the behavior of the
stress as the gas flows through the ``plunging region'' inside of the
marginally stable orbit.  These simulations used the finest
three-dimensional resolution to date, with up to $256\times 64\times
192$ zones in $(R,\phi,z)$, and as many as 70 radial grid zones lying
between the marginally stable orbit and the horizon.  They find that,
in contrast to standard models, the disk does not sharply truncate at
the marginally stable orbit, nor does the stress vanish.  For both
poloidal and toroidal fields the stress actually increases somewhat in
the plunging region.  The nature of the flow shifts from MHD turbulence
to flux freezing, and in doing so drags out and shears the field lines,
increasing the correlation between $B_R$ and $B_\phi$, and hence the
Maxwell stress.  Variability over large spatial scales is the rule
here, over a broad range of time scales.

\begin{figure}[t]
\begin{center}
\includegraphics[width=.9\textwidth]{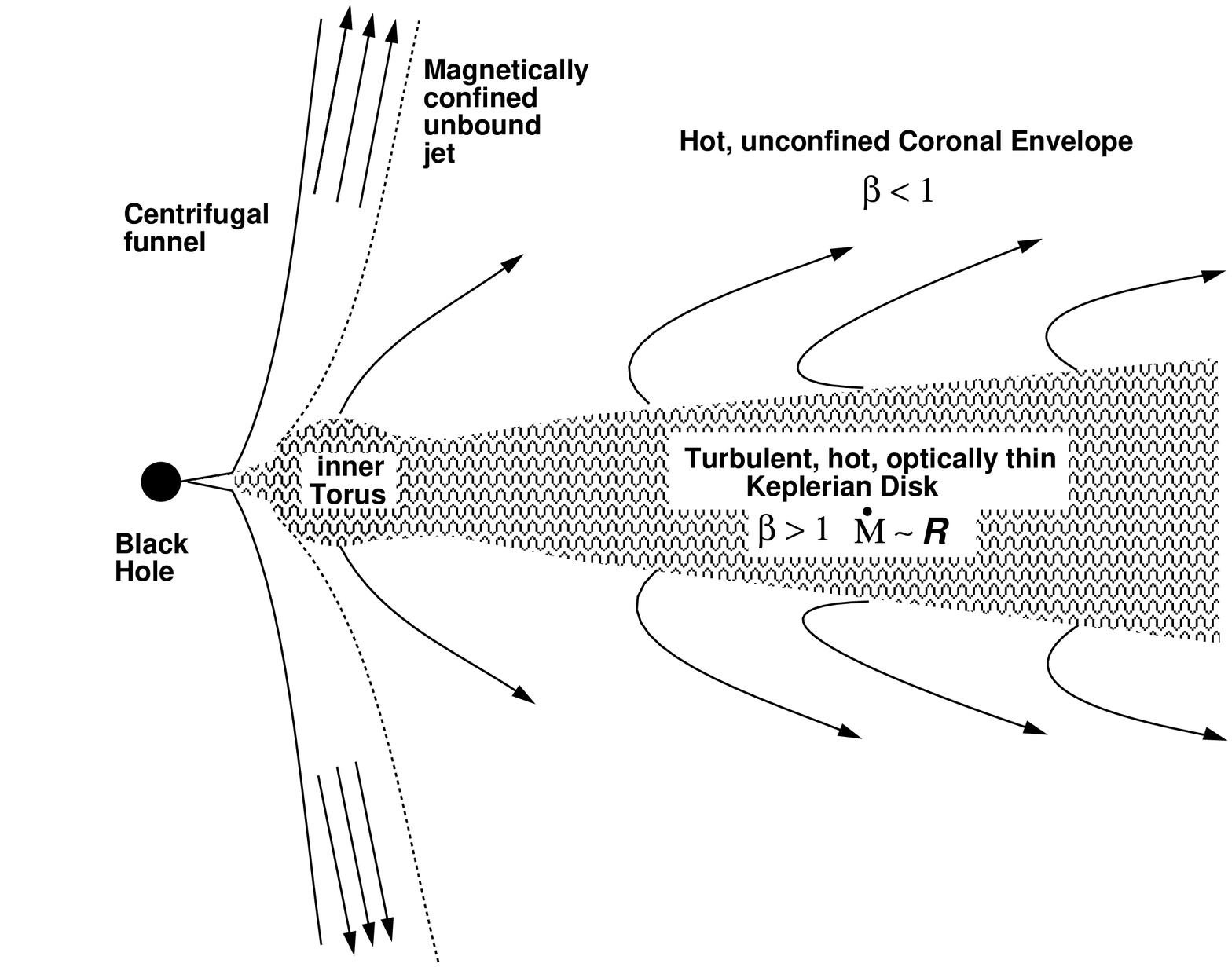}
\end{center}
\caption[]{
A schematic diagram of a nonradiative accretion flow, highlighting its
principal features.  A turbulent, nearly Keplerian gas-dominated hot
disk is surrounded by an active, diffuse, magnetic-dominated coronal
envelope.  Near the marginally stable orbit, the flow thickens into a
small inner torus.  A centrifugally-evacuated funnel lies along the
axis, surrounded by a jet confined by magnetic pressure in the corona.
From Hawley \& Balbus (2002).} \label{eps1} \end{figure}

The general accretion problem occurs over much larger scales, at least
hundreds of gravitational radii.  This global nonradiative MHD
accretion flow, first considered in two dimensions by Stone \& Pringle
(2001), was extended to three dimensions in Hawley, Balbus \& Stone
(2001) and Hawley \& Balbus (2002).  The accretion flow originates with
a constant specific angular momentum torus initially centered at 100
gravitational radii. MHD turbulence ensues and the resulting flow seems
to settle into three well-defined dynamical structures.   The main
accretion is through a hot, thick, rotationally-dominated Keplerian
disk.  The MRI is so efficient at transporting angular momentum, the
transformation from constant angular momentum to Keplerian profile
occurs within a few orbital times at the pressure maximum.  Evidently,
Keplerian profiles are characteristic of warm and cool disks alike.
The total pressure scale height in this disk is comparable to the
vertical size of the initial torus.  Surrounding this disk is a
magnetized corona with vigorous circulation and possibly outflow.  Gas
pressure dominates only near the equator; magnetic pressure is more
important in the surrounding corona.  Finally, a magnetically-confined
jet forms along the centrifugal funnel wall.  Runs with and without
Ohmic heating were performed; very few differences were found.  The
flow was somewhat hotter in the resistive runs, and the turbulence
slightly subdued, but the dynamics remain dominated by MHD turbulence.
There was no evidence of a convective envelope (Abramowicz et al.
2002).

Figure (\ref{eps1}) shows schematically the appearance of the disk,
corona, and jet as they appear in a typical run.  The inner torus is
pressure-thickened, but remains predominately supported by rotation.
It is a highly transitory structure, forming and collapsing over the
course of the simulation.

\begin{figure}[b]
\begin{center}
\includegraphics[width=.8\textwidth]{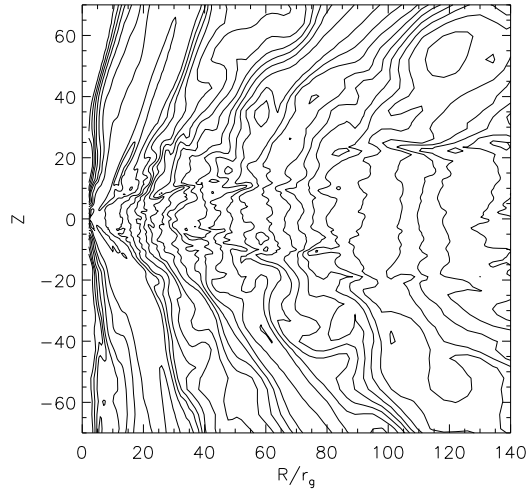}
\end{center}
\caption[]{ Contours of azimuthally-averaged specific angular
momentum.  The disk is associated with the contours stratified nearly
on cylinders.  From Hawley \& Balbus (2002).}

\label{ell}
\end{figure}

Contours of specific angular momentum are shown in figure (\ref{ell}).
The disk emerges sharply in this diagram as the zone of cylindrical
contours.  This appears to be a consequence of the relatively small
magnetic to thermal energy density ratio, and a barotropic equation
of state.  

\begin{figure}[t]
\begin{center}
\includegraphics[width=.9\textwidth]{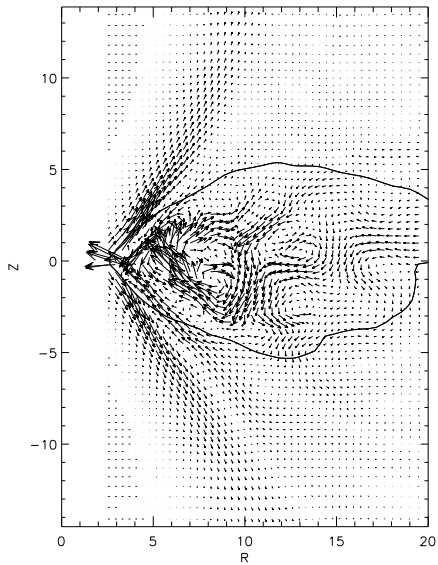}
\end{center}
\caption[]{
Azimuthally-averaged momentum vectors in the within 20
gravitatinal radii of the black hole.
The overlaid contour is density,
showing the shape of the inner torus.  The jet outflow along the
centrifugal barrier
is clearly evident.  The 
magnetic pressure in the corona confines the jet externally.
From Hawley \& Balbus (2002).}
\label{momfig}
\end{figure}

Figure (\ref{momfig}) is a momentum plot of the inner 20 gravitational
radii of the disk, showing the internal dynamical structure of the
inner torus.  The distinct jet structure stands out particularly clearly.

\section{Summary}

Our understanding of accretion phenomena has grown enormously in the
past decade, and the physical process underlying the anomalous
viscosity of accretion disks have been elucidated.  Indeed, ``anomalous
viscosity'' is a misnomer, since magnetic fields do so much more than
fill the role occupied by a  hydrodynamical Navier-Stokes viscosity,
and we should move away from this mode of thought.  Magnetized fluids
are far too subtle for this approach to be successful.

The combination of a magnetic field and outwardly decreasing
differential rotation is highly unstable, as is the combination of a
magnetic field and an upwardly decreasing temperature profile.
Local simulations of Keplerian disks show that this magnetorotational
instability leads to a turbulence-enhanced stress tensor that
transports energy and angular momentum outwards, allowing accretion to
proceed.  The typical dimensionless value of the stress (normalized to a
fiducial pressure) ranges from $5\times 10^{-3}$ to $0.6$ depending
upon field geometry, and is highly variable in both space and time.
Local simulations have become very sophisticated in the class of
problems they are able to investigate.   Work has begun on the local
disk dynamics of radiation-dominated and non-ideal MHD systems.

Fully global three-dimensional MHD simulations are now a reality.
These runs show that initially non-Keplerian angular momentum
distributions rapidly evolve to Keplerian.  Efficient angular momentum
transport builds up rotation in the outer regions in the early stages,
and the gas expands.  Any initial radial pressure gradient is
``inflated'' to zero, and a Keplerian distribution emerges.

Chandra observations have provided compelling evidence that very low
luminosity accretion flows are present in Nature, and such flows are
amenable to numerical investigation.  The most detailed studied to date
(Hawley \& Balbus 2002) reveals a three component structure:  a warm
Keplerian disk, a highly magnetized corona, and an axial jet.  None of
these structures were present in the initial condition, which is a
simple constant angular momentum torus located 100 gravitational radii
from the hole.  Significant fluctuations in all flow quantities are
present, both in time and in space.  There was no indication that
thermal convection was dominating the dynamical flow structure
(Abramowicz, et al. 2002).

The results of global three-dimensional MHD numerical simulations
comprise a vast and extremely useful data base, a true numerical
laboratory.  At the time of this writing, relatively little has been
done to translate the formal numerical flows into photons impinging
upon instruments.  The exploitation of this largely untapped resource
should provide advances as singular as those of the last extraordinary
decade.

\section*{Acknowldgements}

Our research is supported by NASA grants NAG-10655, NAG5-9266, and NSF
grant AST-0070979.  The authors are most grateful to the editors E.
Falgarone and T. Passot for the their patience and forbearance.

\end{document}